\documentclass[11pt]{article}
\usepackage{xspace}
\usepackage{graphicx}
\usepackage{amsmath}
\usepackage{amssymb}
\usepackage{color}
\usepackage{hyperref}

\usepackage{tikz}
\usetikzlibrary{decorations.pathmorphing,decorations.markings}

\definecolor{Red}{rgb}{1,0,0}
\definecolor{Green}{rgb}{0,1,0}
\definecolor{Blue}{rgb}{0,0,1}
\definecolor{Black}{rgb}{0,0,0}

\def\beq{\begin{equation}}
\def\eeq#1{\label{#1}\end{equation}}
\def\eeqn{\end{equation}}

\def\beqa{\begin{eqnarray}}
\def\eeqa#1{\label{#1}\end{eqnarray}}
\def\eeqa{\end{eqnarray}}

\let\bar=\overbar

\def\Dslash{\not{\hbox{\kern-4pt $D$}}}
\def\dslash{\not{\hbox{\kern-2pt $\del$}}}

\def\msb{{\bar{\ssstyle M \kern -1pt S}}}

\def\babar{\mbox{\slshape B\kern-0.1em{\small A}\kern-0.1em
    B\kern-0.1em{\small A\kern-0.2em R}}}

\def\beq{\begin{equation}}
\def\eeq{\end{equation}}

\tikzset{
boson/.style={decorate, decoration={snake,amplitude=2pt, segment length=5pt}, draw=black},
scalar/.style={dashed, draw=black},
particle/.style={draw=black, postaction={decorate}, decoration={markings,mark=at position .5 with {\arrow[draw=black]{>}}}},
antiparticle/.style={draw=black, postaction={decorate}, decoration={markings,mark=at position .5 with {\arrow[draw=black]{<}}}},
gluon/.style={decorate, draw=black, decoration={coil,amplitude=4pt, segment length=5pt}},
sfermion/.style={dashed, draw=black, postaction={decorate}, decoration={markings,mark=at position .5 with {\arrow[draw=black]{>}}}},
antisfermion/.style={dashed, draw=black, postaction={decorate}, decoration={markings,mark=at position .5 with {\arrow[draw=black]{<}}}},
gaugino/.style={draw=black},
}

\def\Title#1{\begin{center} {\Large {\bf #1} } \end{center}}

\begin{document}

\Title{Constraints on dark forces from the $B$~factories 
       and low-energy experiments}

\bigskip\bigskip


\begin{raggedright}  

Abner Soffer\index{Soffer, A.}, {\it Tel Aviv University}\\

\begin{center}\emph{On the behalf of the \babar\ Collaboration.}\end{center}
\bigskip
\end{raggedright}

\begin{abstract}
The idea that dark-matter interactions with Standard-Model particles
may be mediated by new bosons with masses in the MeV-to-GeV range took
off several years ago.  Constraints on such models were soon
calculated based on older measurements. Subsequently, active
collaborations conducted dedicated searches for these bosons, and new
experiments were planned to improve the search sensitivity. I review
the basic models that predict dark vectors and dark Higgs bosons in
this mass range, the constraints from electron-positron colliders,
fixed-target experiments, and hadron colliders, and comment on the
sensitivities of future experiments.
\end{abstract}

{\small
\begin{flushleft}
\emph{To appear in the proceedings of the Interplay between Particle and Astroparticle Physics workshop, 18 -- 22 August, 2014, held at Queen Mary University of London, UK.}
\end{flushleft}
}

\section{Dark forces}
The colorful term ``dark forces'' refers to interactions involving
dark-matter particles, particularly to the extent that they serve as
``portals'' between the Standard Model (SM) particles and those of the
dark-matter sector (DS). Recently, scenarios in which such
interactions are mediated by GeV-scale particles have generated a
great deal of interest. Such a model was proposed in
Ref.~\cite{nima-DM-theory} in order to explain chiefly the rise in
the cosmic-ray positron fraction with energy, starting around 10~GeV,
as seen by PAMELA~\cite{pamela} and later confirmed with high
precision by AMS-02~\cite{ams2}. This rise is also consistent with
secondary positron production due to collisions of primary cosmic rays
with interstellar gas and dust. However, the idea that it may partly
be due to physics beyond the Standard Model has proven almost
revolutionary: it has motivated much theoretical and experimental work
on new, GeV-scale states, including the construction of new
experiments.

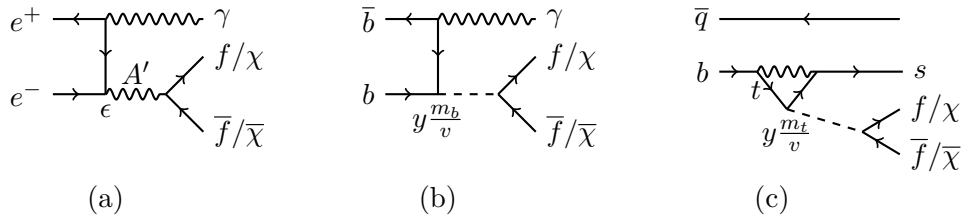
\begin{figure}[!ht]
\begin{center}
%

\input diag-defs.tex

\begin{tabular}{ccc}
\begin{tikzpicture}[thick,scale=1.0]
\draw (-0.5,-1.9) node[]{(a)};
\draw[antiparticle] (-1.2,0.5) node[black,left]{$e^+$} -- (-0.5,0.5);
\draw[particle] (-1.2,-0.5) node[black,left] {$e^-$} -- (-0.5,-0.5);
\draw[particle] (-0.5,0.5) -- (-0.5,-0.5) node[black, below] {$\epsilon$};
\draw[boson] (-0.5,0.5) -- (0.8,0.5) node[black, right]{$\gamma$} ;
\draw[boson] (-0.5,-0.5) -- node[black, above]{$A'$} (0.3,-0.5) ;
\draw[particle] (0.3,-0.5) -- (0.8, 0) node[black,right] {$f/\chi$} ;
\draw[antiparticle] (0.3,-0.5) -- (0.8, -1) node[black,right] {$\bar f/\bar\chi$} ;
\end{tikzpicture}

& 
\hspace{0.5cm}

\begin{tikzpicture}[thick,scale=1.0]
\draw (-0.5,-1.9) node[]{(b)};
\draw[antiparticle] (-1.2,0.5) node[black,left]{$\bar b$} -- (-0.5,0.5);
\draw[particle] (-1.2,-0.5) node[black,left] {$b$} -- (-0.5,-0.5);
\draw[particle] (-0.5,0.5) -- (-0.5,-0.5) node[black, below] {$y{m_b\over v}$};
\draw[boson] (-0.5,0.5) -- (0.8,0.5) node[black, right]{$\gamma$} ;
\draw[scalar] (-0.5,-0.5) -- 
                            (0.3,-0.5) ;
\draw[particle] (0.3,-0.5) -- (0.8, 0) node[black,right] {$f/\chi$} ;
\draw[antiparticle] (0.3,-0.5) -- (0.8, -1) node[black,right] {$\bar f/\bar\chi$} ;
\end{tikzpicture}

& 
\hspace{0.5cm}

\begin{tikzpicture}[thick,scale=1.0]
\draw (-0.5,-1.9) node[]{(c)};
\draw[antiparticle] (-1.2,0.5) node[black,left]{$\bar q$} -- (1.2,0.5);
\draw[particle] (-1.2,-0.2) node[black,left] {$b$} -- (-0.7,-0.2);
\draw[particle] (0.1,-0.2) -- (1.24,-0.2) node[black,right] {$s$} ;
\draw[boson] (-0.7,-0.2) -- (0.1,-0.2);
\draw[particle] (-0.7,-0.2) -- node[black,left] {$t$} (-0.3,-0.7) node[black,below] {$y {m_t\over v}$} ;
\draw[particle] (-0.3,-0.7)  -- (0.1,-0.2);
\draw[scalar] (-0.3,-0.7) -- (0.7,-1.0);
\draw[particle] (0.7,-1.0) -- (1.2, -0.7) node[black,right] {$f/\chi$} ;
\draw[antiparticle] (0.7,-1.0) -- (1.2, -1.3) node[black,right] {$\bar f/\bar\chi$} ;
\end{tikzpicture}

\begin{tikzpicture}[thick,scale=1.0]
\end{tikzpicture}

\end{tabular}

\caption{Feynman diagrams for (a) dark-photon production in $e^+e^-$
  collisions, (b) dark-Higgs production in $\Upsilon$ decay, and (c)
  dark-Higgs production in penguin $B$ decay.  The dark photon $A'$ or
  dark Higgs $\phi$ is shown decaying into a pair of SM
  fermions $f\bar f$ or invisible dark-sector fermions $\chi\bar\chi$.}
\label{diags}
\end{center}
\end{figure}

We describe here two types of portals. In the vector portal, one
postulates the existence of a $U(1)$ gauge interaction in the dark
sector, which mixes with the SM $U(1)_Y$. After electroweak symmetry
breaking, the effective Lagrangian mixes the
associated dark photon $A'$ with the SM photon:
\beq
{\cal L}_{\rm eff} = {\cal L}_{\rm SM} - {1 \over 4} F'_{\mu\nu} F'^{\mu\nu}
 + {m_{A'}^2 \over 2} A'_\mu A'^\mu 
 - {\epsilon \over 2} F'_{\mu\nu} F^{\mu\nu},
\eeq
where $F'_{\mu\nu}$ is the dark photon field, $\epsilon$ is the
effective mixing parameter, and $m_{A'}$ is the dark photon mass,
which may be generated by the breaking of a larger symmetry.
Phenomenologically, a dark photon may be created in electromagnetic
processes, replacing a virtual SM photon, and may then decay back
into a pair of SM fermions $f\bar f$ or dark-sector fermions (WIMPs)
$\chi\bar\chi$. In $e^+e^-$ collisions, the
relevant Feynman diagram is shown in Fig.~\ref{diags}(a).

The Higgs portal features a light scalar $\phi$, which mixes slightly
with the SM Higgs, and therefore has mass-proportional couplings to
the SM fermions. The effective Lagrangian may be written
as~\cite{Schmidt-Hoberg:2013hba}
\beq
{\cal L}_{\rm eff} = {\cal L}_{\rm SM} - y {m_f \over v}\phi \bar f f
                       - {1\over 2} \kappa \phi \bar\chi\chi,
\label{eq:higgs-mixing-lang}
\eeq
where $y$ is the effective scalar-mixing parameter, and $\kappa$ is
the dark-Higgs coupling to the WIMP. The
$\phi\bar f f$ term enables creation of the dark Higgs in radiative
decays of the narrow $\Upsilon(nS)$ resonances (where $n=1,2,3$),
shown in Fig.~\ref{diags}(b). Production in radiative decays of the
$J/\psi$ are also interesting, although they are suppressed due to the
small charm-quark mass.
Another possibility for production of the dark Higgs is in penguin
$B$-meson decays, shown in Fig.~\ref{diags}(c). These have two
advantages over $\Upsilon$ decays: the first is that $B$ mesons are
many orders of magnitude narrower than the $\Upsilon$ states, and the
second is the large coupling of the dark Higgs to the top quark in the
penguin loop. On the other hand, penguin $B$ decays have a very small
branching fraction compared with radiative $\Upsilon$
decays. Furthermore, production in $B$ decays is limited to dark-Higgs
masses of $ m_\phi \lesssim 4.5$~GeV.

\section{$B$ factories and other dark-forces facilities}
Electron-positron $B$~factories are well suited for searching for new
physics at the GeV scale, mainly due to their large data samples.
Together, \babar~\cite{TheBABAR:2013jta,Aubert:2001tu}
and Belle~\cite{Abashian:2000cg} have collected about
1.6~fb$^{-1}$~\cite{lumi-paper} at and around the $\Upsilon$
resonances. This large sample, plus the sizeable
$e^+e^-\to\gamma\gamma$ cross section of about 3~nb at $B$-factory
energies, give an idea of the $\epsilon$ sensitivity of these
experiments.

Fixed-target experiments typically have much larger integrated
luminosities and lower center-of-mass energies than collider
experiments. As a result, they are sentivive to lower values of
$\epsilon$ at lower regions of $m_{A'}$.

The Higgs-portal sensitivity of the $B$~factories stems from their
large sample of $B$ mesons, pair-produced in $\Upsilon(4S)$ decays,
as well as samples of the narrow $\Upsilon(1S,2S,3S)$ resonances.
\babar\ has collected
$(470.9 \pm 2.8)\times 10^6$ $\Upsilon(4S)$ mesons~\cite{McGregor:2008ek}, 
$(121.8 \pm 1.2)\times 10^6$ $\Upsilon(3S)$ mesonss, 
and $(98.6 \pm 0.9)\times 10^6$ $\Upsilon(2S)$ mesons. 
The numbers for Belle are
$657\times 10^6$ $\Upsilon(4S)$, 
$3\times 10^6$ $\Upsilon(3S)$, 
$25\times 10^6$ $\Upsilon(2S)$, 
and $6\times 10^6$ $\Upsilon(1S)$.

Hadron-collider experiments produce $B$ mesons with a much larger
cross section than $e^+e^-$ colliders. However, they typically require
muons to trigger on, and are therefore sensitive mostly to $B\to
K\mu^+\mu^-$.

\section{Vector-portal constraints}
\begin{figure}[!ht]
\begin{center}
\includegraphics[width=1.0\columnwidth]{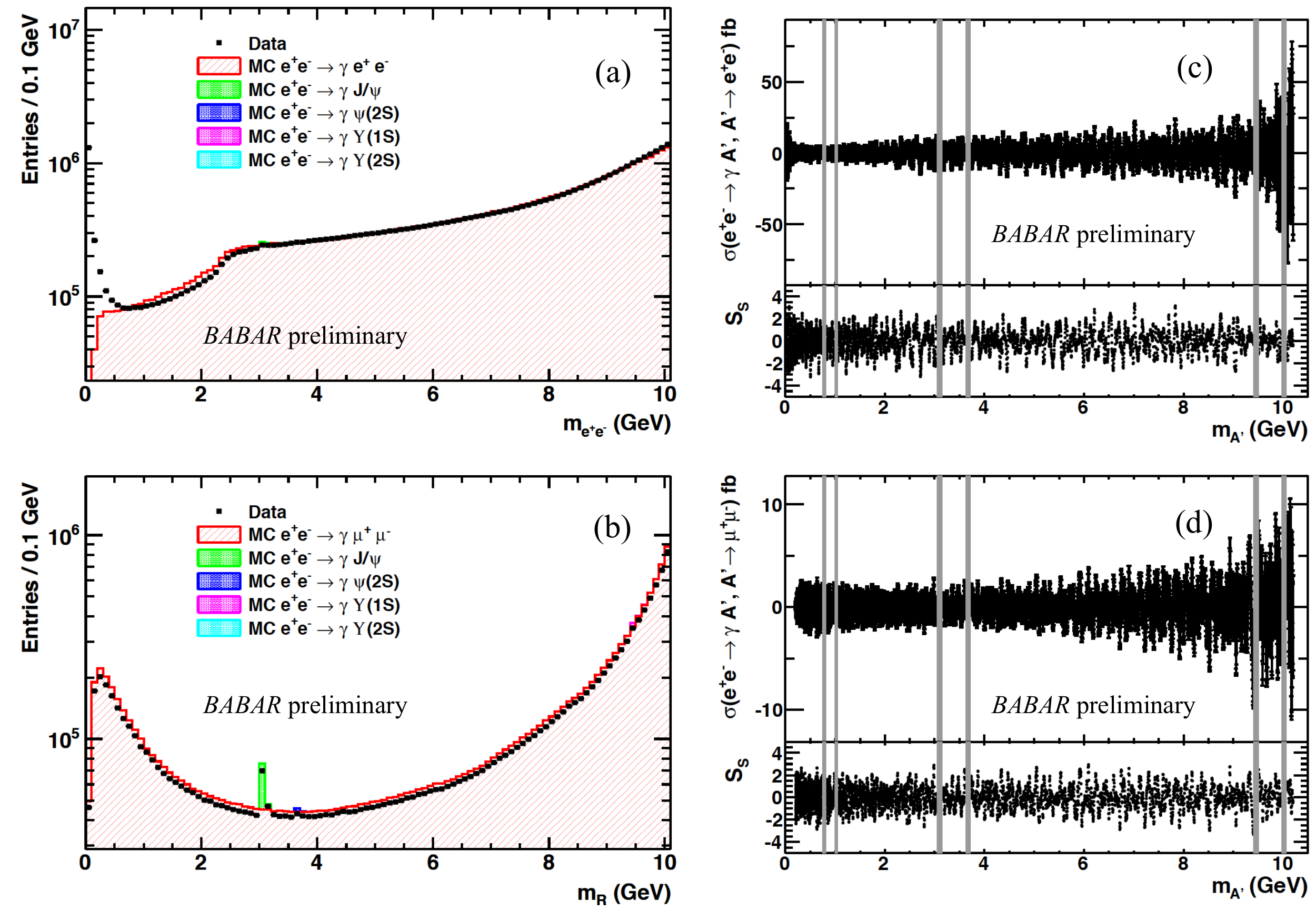}
\caption{(a) Distributions of the invariant mass for $e^+e^-$ pairs
  and (b) the reduced mass $m_R = \sqrt{m_{\mu\mu}^2 - 4m_\mu^2}$ for
  $\mu^+\mu^-$ pairs, and (c,d) the respective
  signal yields and significances as functions of mass in the
  \babar\ dark-photon search{\protect{~\cite{Lees:2014xha}}}.  }
\label{babar-DP-distribs}
\end{center}
\end{figure}

\babar\ has searched for a dark photon produced via the diagram in
Fig.~\ref{diags}(a), in the final states $A'\to e^+e^-$ and
$\mu^+\mu^-$~\cite{Lees:2014xha}.  Figs.~\ref{babar-DP-distribs}(a)
and~\ref{babar-DP-distribs}(b) show the invariant-mass distribution of
the $e^+e^-$ pairs and the reduced mass $m_R = \sqrt{m_{\mu\mu}^2 -
  4m_\mu^2}$, which is better modeled at low values, for $\mu^+\mu^-$
pairs. Several known SM resonances are visible on top of a smooth
continuum background from radiative Bhabha scattering and dimuon
events. Discrepancies between the data and the Monte Carlo predictions
are visible, particularly in the low $m_{e^+e^-}$ region, for which
the Monte-Carlo generator is not designed.  However, the analysis does
not rely on the generator.  \babar\ search for a signal by fitting for
a single signal peak, which is moved in small mass steps. In each fit,
a mass region at least 20 times wider than the signal resolution is
used, and the background shape is taken to be a third- or fourth-order
polynomial. The signal yields and significances as functions of mass
are shown in Figs.~\ref{babar-DP-distribs}(c)
and~\ref{babar-DP-distribs}(d) for the two modes, respectively.

\begin{figure}[!ht]
\begin{center}
\includegraphics[width=0.8\columnwidth]{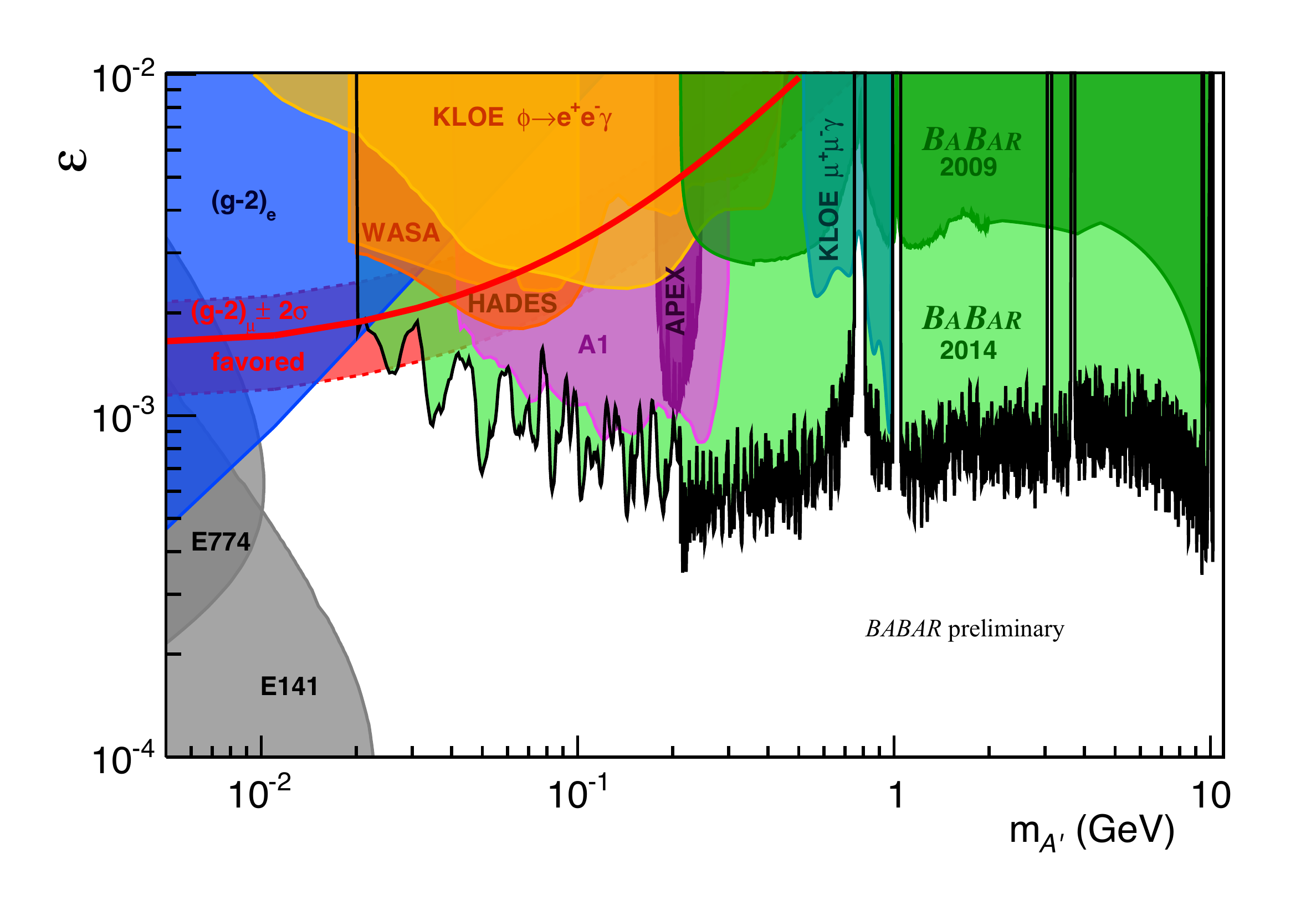}
\includegraphics[width=0.75\columnwidth]{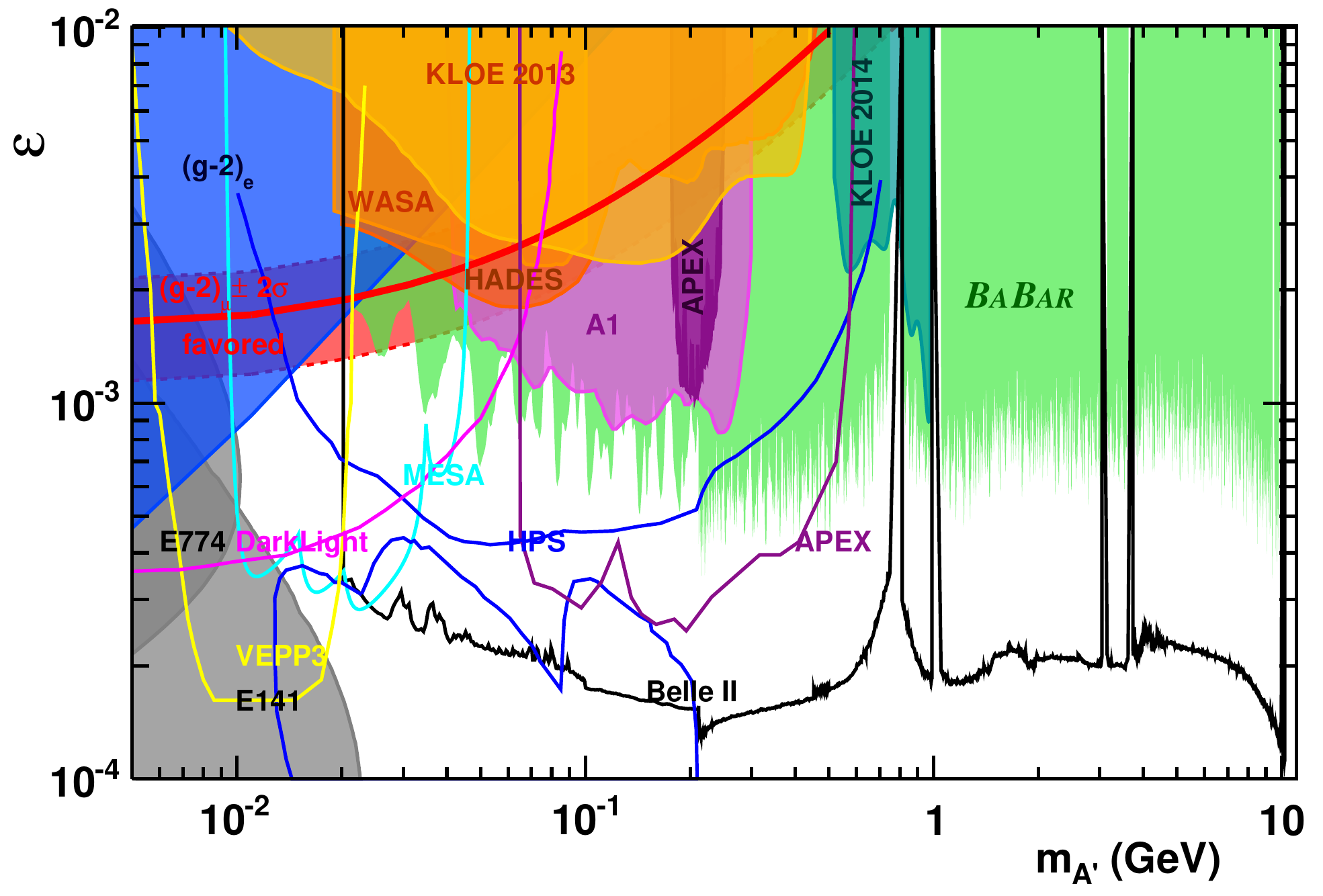}
\caption{Top: Upper limits on the dark-photon mixing parameter
  $\epsilon$ vs. its mass $m_{A'}$ from various experiments. See text
  for details. Bottom: Current limits with projected sensitivities of
  future experiments.  }
\label{babar-DP-limits}
\end{center}
\end{figure}

\babar\ did not observe a significant signal, and set 90\%
confidence-level (CL) upper limits on the dark-photon mixing parameter
$\epsilon$ vs. its mass $m_{A'}$. These preliminary constraints are labeled
``\babar\ 2014'' in Fig.~\ref{babar-DP-limits}.
Also shown in Fig.~\ref{babar-DP-limits} are several other results,
which at high mass are less sensitive.  These are a reinterpretation
of a \babar\ search for a light Higgs (labeled ``\babar\ 2009''. See
Sec.~\ref{sec:higgs-search})~\cite{Aubert:2009cp};
dark-photon searches by KLOE using $e^+e^-\to
  \mu^+\mu^-\gamma$~\cite{Babusci:2014sta} and $\phi(1020)\to\eta
  e^+e^-$~\cite{Babusci:2012cr};
searches for $A'\to e^+e^-$ by the electron-nucleus fixed-target
  experiments APEX~\cite{Abrahamyan:2011gv} and
  A1~\cite{Merkel:2014avp};
a $A'\to e^+e^-$ search by the proton fixed-target experiment
  HADES~\cite{Agakishiev:2013fwl};
and a search by WASA using $\pi^0\to r^+e^-\gamma$~\cite{Adlarson:2013eza}.
For small $m_{A'}$ values, limits have been
obtained~\cite{Andreas:2012mt} from older electron-beam-dump
experiments, of which only constraints from E774~\cite{Bross:1989mp} and
E141~\cite{Riordan:1987aw} are shown in Fig.~\ref{babar-DP-limits}, as
well as proton-beam-dump experiments and the cooling rate of supernova
SN1978A (not shown)~\cite{Essig:2013lka}.
Also shown in Fig.~\ref{babar-DP-limits} are the constraints
from measurement of the electron magnetic moment~\cite{Endo:2012hp} and the
band of favored $\epsilon$ vs. $m_{A'}$ values obtained if one
attributes to the dark photon the discrepancy between the measured and
SM-calculated values of the muon magnetic
moment~\cite{Pospelov:2008zw}.  This favored band is now almost
completely excluded by the results shown, as well as by a new
$e^+e^-\to e^+e^-\gamma$ search from KLOE~\cite{KLOE-ICHEP}, not shown
in Fig.~\ref{babar-DP-limits}.

Tighter constraints will be obtained from future measurements.
Between them, the DarkLight, APEX, and HPS experiments~\cite{Boyce:2012ym}
at Jefferson Lab will be able to exclude $\epsilon$ values above about
$3\times10^{-4}$ for a broad range of $m_{A'}$ between $10^{-6}$~eV
and 300~MeV.
HPS will also search for displaced vertices formed by decays of
long-lived dark photons, with sensitivity to $\epsilon$ between about
$10^{-5}$ and $10^{-4}$ in the approximate mass range $30<m_{A'}<200$~MeV.
A search by Belle should yield limits similar to those of \babar,
given that the analysis is background-dominated, so that the
$\epsilon$ sensitivity scales as the fourth root of the integrated
luminosity.  Belle-II, however, will have 100 times more
integrated luminosity than \babar, and factors of 2 better trigger
efficiency and mass resolution, thus reaching roughly 6-fold
tighter limits on $\epsilon$.
The predicted sensitivities of these future experiments are also shown in 
Fig.~\ref{babar-DP-limits}.

\begin{figure}[!ht]
\begin{center}
\includegraphics[width=1.0\columnwidth]{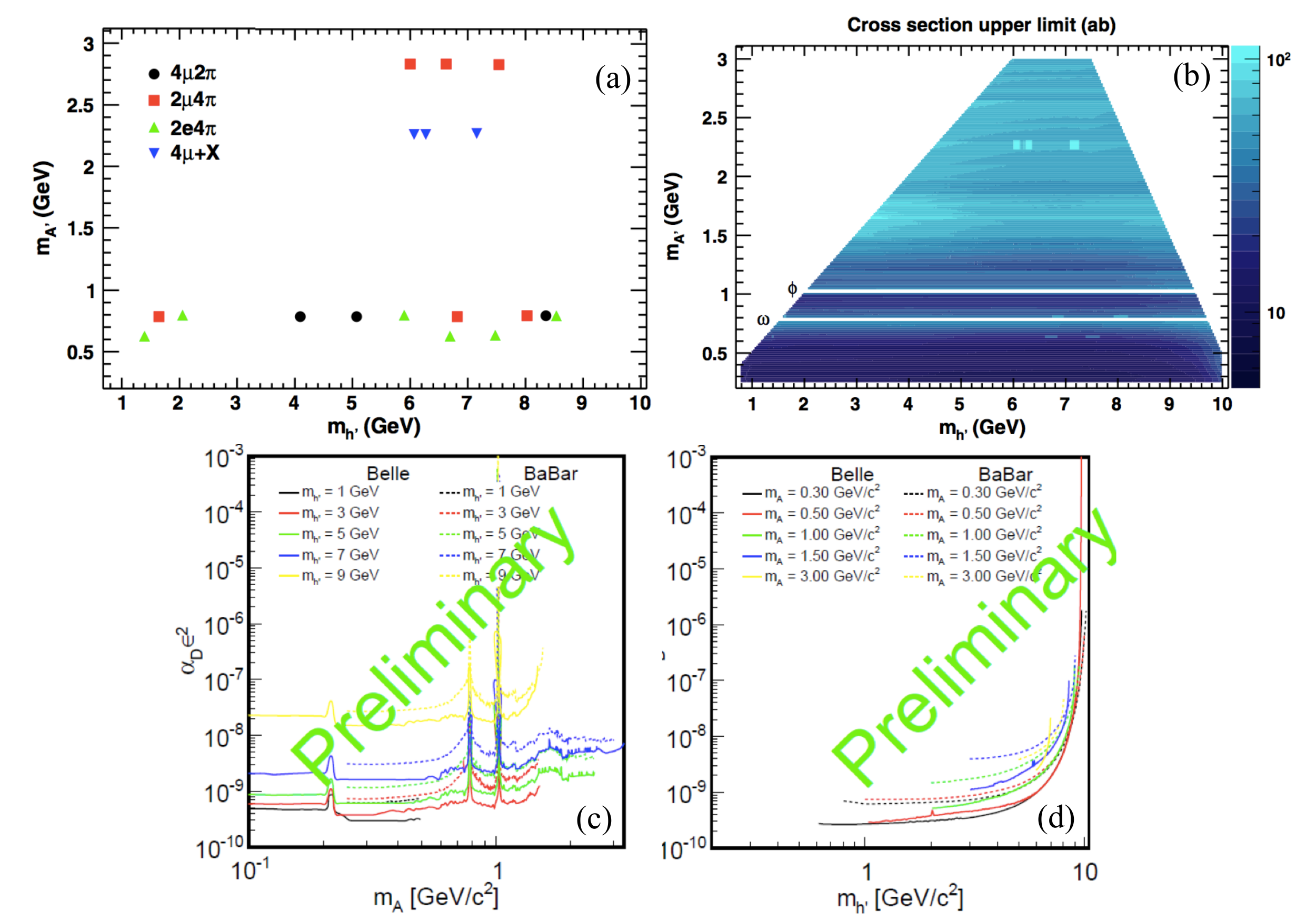}
\caption{(a) The dark-photon mass $m_{A'}$ vs. dark-Higgs mass
  $m_{h'}$ for the three possible $h'\to A'A'$ assignments for each of
  the 6 events seen by \babar~{\protect{\cite{Lees:2012ra}}} in the
  search for $e^+e^-\to A'A'A'$. (b) The resulting 90\% CL limits on
  the production cross section.  (c,d) The extracted limits on
  $\epsilon^2$ times $\alpha_D=g_D^2/4\pi$, where $g_D$ is the
  dark-sector gauge coupling, as a function of $m_{A'}$ and $m_{h'}$,
  from Belle~{\protect{\cite{Belle-3A'}}} (solid lines) and
  \babar~{\protect{\cite{Lees:2012ra}}} (dashed lines).  }
\label{fig:Higgsstrahlung}
\end{center}
\end{figure}

\babar\ and Belle have also searched for a dark photon in the
dark-Higgsstrahlung scenario. In this case, the $A'$ radiates an
on-shell dark Higgs, which decays into two on-shell dark vectors.  The
resulting upper limits from \babar~\cite{Lees:2012ra} and preliminary
limits from Belle~\cite{Belle-3A'} are shown in
Fig.~\ref{fig:Higgsstrahlung}. We note that the final state of three
dark photons with the same mass has very little background, so that
the limits improve approximately linearly with luminosity, implying a
two-orders-of-magnitude improvement at Belle-II.

\section{Higgs-portal constraints}
\label{sec:higgs-search}

Although experiments have not searched directly for dark-Higgs states,
they have studied processes with the same final-state topology, which
can be used to set limits on dark-Higgs production. 

\subsection{Searches in $\Upsilon$ decays}
\label{sec:upsilon}

\babar\ has searched for the light, CP-odd Higgs $A^0$ of the NMSSM
scenario~\cite{Maniatis:2009re,Dermisek:2010mg} using two methods.  In
the first method, one looks for production of $A^0$ in the radiative
decay $\Upsilon(2S, 3S)\to A^0\gamma$ (Fig.~\ref{diags}(b)), with
subsequent $A^0$ decay into $\mu^+\mu^-$~\cite{Aubert:2009cp},
$\tau^+\tau^-$~\cite{Aubert:2009cka}, hadrons~\cite{Lees:2011wb}, or
invisible particles~\cite{Aubert:2008as}.
This method has also been used by CLEO in radiative $\Upsilon(1S)$
decays, using a data sample of $1~{\rm fb}^{-1}$,with the $A^0$ final
states $\mu^+\mu^-$ and $e^+e^-$~\cite{Love:2008aa}.
The BES-III experiment has conducted a similar search in radiative $J/\psi$
decays~\cite{Ablikim:2011es}.

In the second method, the decay
$\Upsilon(2S)\to\Upsilon(1S)\pi^+\pi^-$ is identified by
reconstructing just the two pions and requiring the squared recoil
mass $(p_{e^+e^-} - p_{\pi^+\pi^-})^2$ to be consistent with
production of an $\Upsilon(1S)$. This essentially eliminates the
non-$\Upsilon$ background, so that despite the relatively small
branching fraction ${\cal B}(\Upsilon(2S)\to\Upsilon(1S)\pi^+\pi^-) =
0.1785 \pm 0.0026$~\cite{Beringer:1900zz}, the two methods have 
comparable sensitivities.
\babar\ has used this method to search for $A^0\to
\mu^+\mu^-$~\cite{Lees:2012iw}, $\tau^+\tau^-$~\cite{Lees:2012te},
hadrons~\cite{Lees:2013vuj}, or invisible
particles~\cite{delAmoSanchez:2010ac}.

\begin{figure}[!ht]
\begin{center}
\includegraphics[width=1.0\columnwidth]{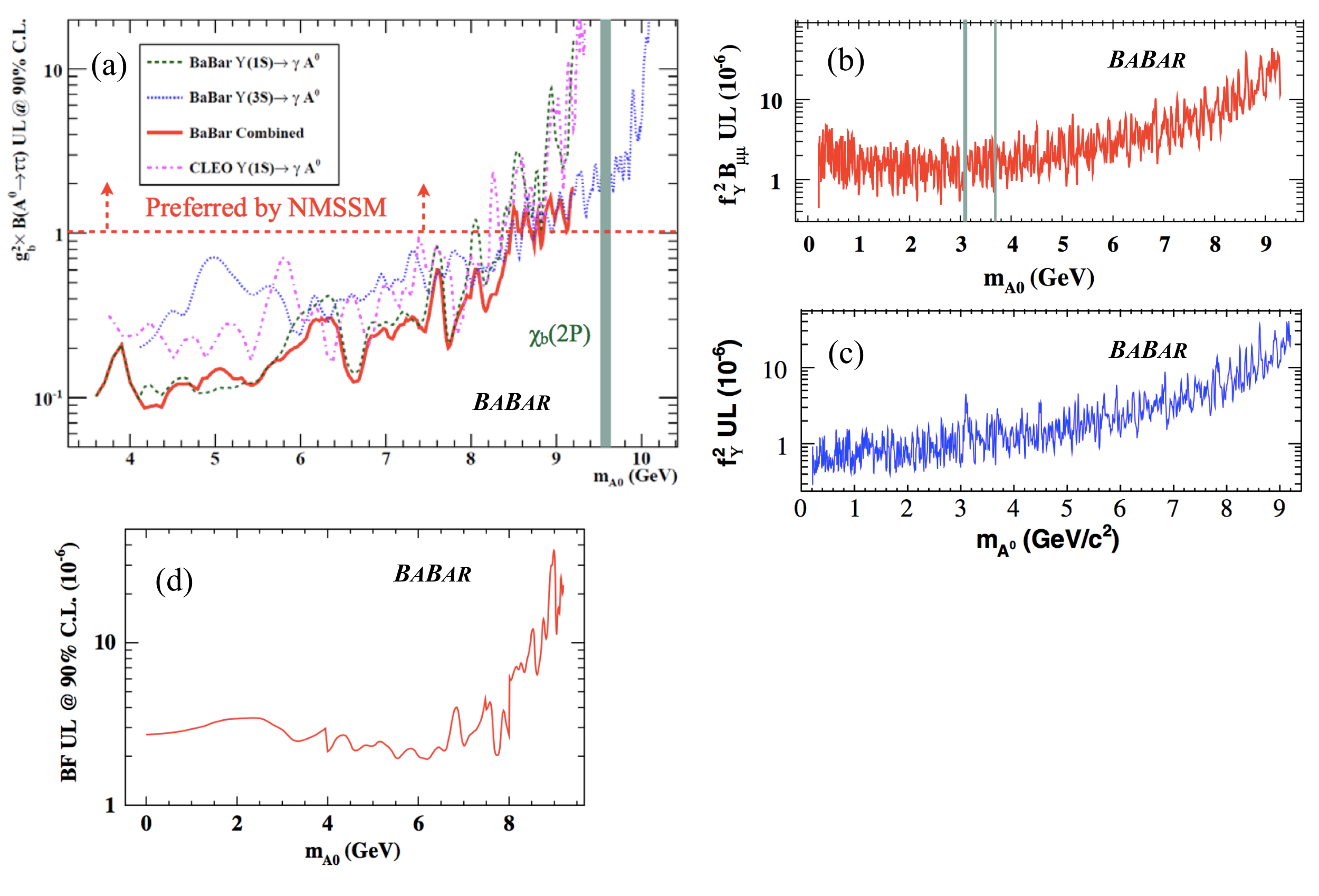}
\caption{\label{fig:A0-leps-invis} \babar\ 90\% CL upper limits, 
  as functions of the light-Higgs mass $m_{A^0}$, on
  (a) $g_b^2 {\cal B}(A^0\to \tau^+\tau^-)$~{\protect{\cite{Aubert:2009cka,Lees:2012te}}} (limits from CLEO~{\protect{\cite{Love:2008aa}}} are also shown),
  on $f_\Upsilon {\cal B}(A^0\to\mu^+\mu^-)$ from 
  (b) $\Upsilon(2S,3S)$ decays~{\protect{\cite{Aubert:2009cp}}}  and 
  (c) $\Upsilon(1S)$ decays~{\protect{\cite{Lees:2012iw}}}, 
  and (d) on ${\cal B}(\Upsilon(1S)\to A^0\gamma) 
  {\cal B}(A^0\to{\rm invisible})$~{\protect{\cite{delAmoSanchez:2010ac}}}.
  (See Eq.~(\protect{\ref{eq:f-and-g}}) for 
  the definitions of $g_b$ and $f_\Upsilon$.)
}
\end{center}
\end{figure}

\begin{figure}[!ht]
\begin{center}
\includegraphics[width=1.0\columnwidth]{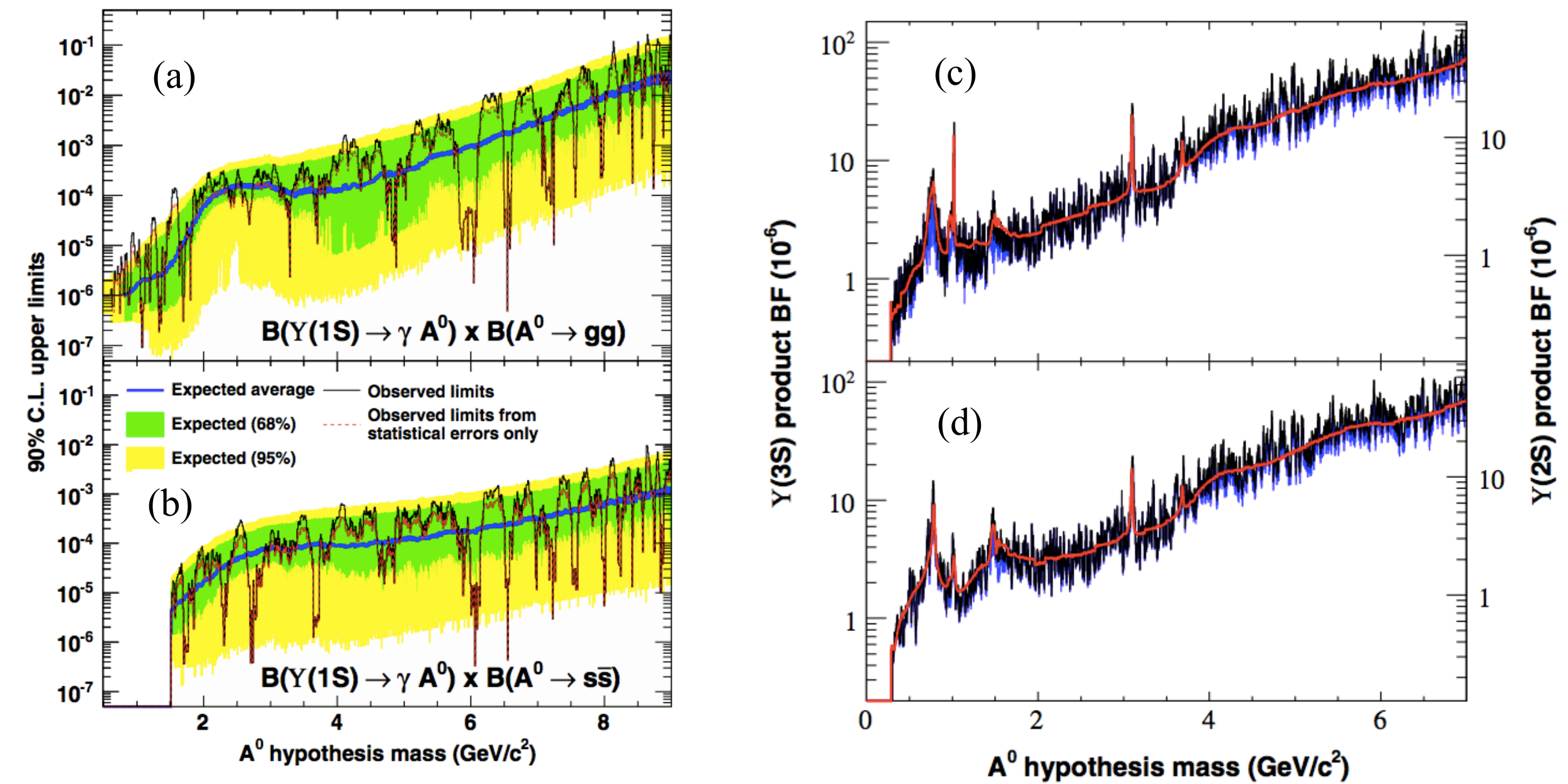}
\caption{\label{fig:A0-hadrons} \babar\ 90\% CL upper limits,
  as functions of the light-Higgs mass $m_{A^0}$, on
  ${\cal B}(\Upsilon(nS)\to A^0\gamma){\cal B}(A^0\to h)$~{\protect{\cite{Lees:2011wb, Lees:2013vuj}}}, where 
  the hadronic final state $h$ is either (a) $gg$, (b) $s\bar s$,
  (c) all hadrons in the event, or (d) all hadrons in the event
  excluding the CP-even states $K^+K^-$ and $\pi^+\pi^-$.
  The $\Upsilon$ states are (a,b) $n=1$ or (c,d) $n=3$ (left vertical axis)
  or $n=2$ (right vertical axis, not including a phase-space correction
  of at most 3.5\%).
}
\end{center}
\end{figure}

A summary of the results is shown in Fig.~\ref{fig:A0-leps-invis} for
$A^0$ decays to leptons and invisible particles, and in
Fig.~\ref{fig:A0-hadrons} for decays to hadronic final states.  The
results are presented as limits on the relevant $A^0$ branching
fraction times either ${\cal B}(\Upsilon(nS)\to A^0\gamma)$ or the
squared couplings $f_\Upsilon$ and $g_b$, defined from
\beqa
{{\cal B}(\Upsilon(nS)\to \gamma A^0) 
   \over {\cal B}(\Upsilon(nS)\to \ell^+\ell^-)}
 &=& {f_\Upsilon^2 \over 2 \pi\alpha} 
      \left(1 - {m_{A^0}^2 \over m_{\Upsilon(nS)}^2}\right), \nonumber\\
f_{\Upsilon}^2 &=& \sqrt{2} g_b^2 G_F m_b^2 F_{\rm QCD},
\label{eq:f-and-g}
\eeqa
where $F_{\rm QCD}$ includes QCD and relativistic corrections of up to
30\% to the $\Upsilon(nS)$ branching fractions.

\subsection{Searches in penguin $B$ decays}
\begin{figure}[!ht]
\begin{center}
\includegraphics[width=1.0\columnwidth]{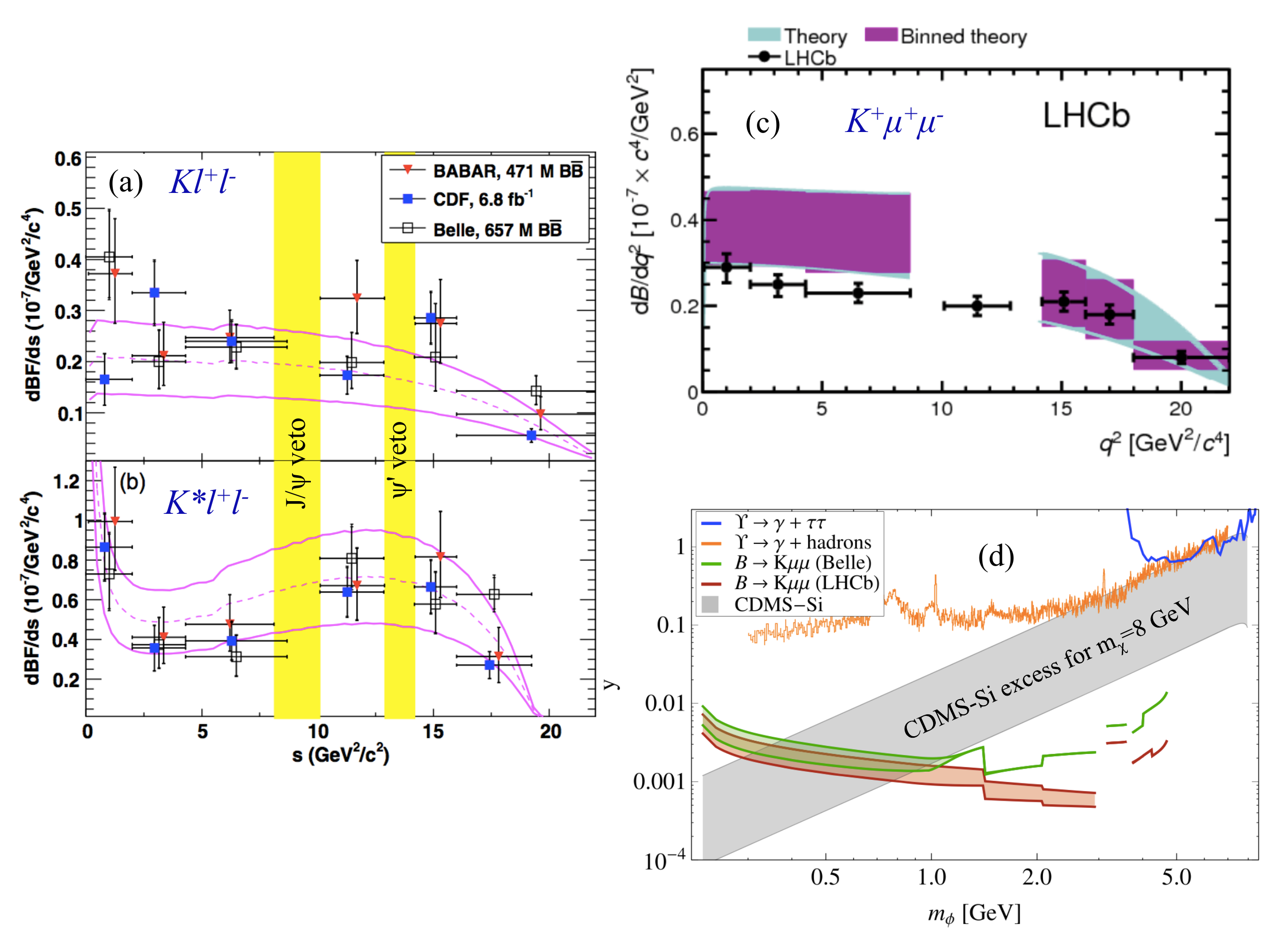}
\caption{\label{fig:BtoKll} The $s=q^2=m(\ell^+\ell^-)$ spectra in
  $B\to K^{(*)}\ell^+\ell^-$ meausred by (a,b)
  CDF~\protect{\cite{Aaltonen:2011qs}}
  Belle~\protect{\cite{Adachi:2008sk}},
  \babar~\protect{\cite{Lees:2012tva}}, and (c)
  LHCb~\protect{\cite{Aaij:2012vr}}.  For CDF and LHCb,
  $\ell^+\ell^-=\mu^+\mu^-$ only, while Belle and \babar\ results
  include also the $e^+e^-$ mode.  (d)
  Limits~\protect{\cite{Schmidt-Hoberg:2013hba}} on the coupling
  parameter $y$ of Eq.~(\protect{\ref{eq:higgs-mixing-lang}}) from
  some of these results (Belle and LHCb) in comparison to some of
  those from $\Upsilon$ decays (see Sec.~\protect{\ref{sec:upsilon}})
  and to results from CDMS~\protect{\cite{Agnese:2013rvf}}
  (interpreted as signal), assuming a WIMP mass of $m_\chi=8$~GeV.  }
\end{center}
\end{figure}

Limits on light-Higgs parameters have also been extracted from studies
of electroweak penguin $B$ decays (Fig.~\ref{diags}(c)).  The dilepton
invariant-mass-squared spectra for $B\to K^{(*)}\ell^+\ell^-$ decays
measured by CDF~\cite{Aaltonen:2011qs}, Belle~\cite{Adachi:2008sk},
\babar~\cite{Lees:2012tva}, and LHCb~\cite{Aaij:2012vr} are shown in
Fig.~\ref{fig:BtoKll}(a,b,c). Fig.~\ref{fig:BtoKll}(d) shows
limits~\cite{Schmidt-Hoberg:2013hba} on the coupling parameter $y$ of
Eq.~(\ref{eq:higgs-mixing-lang}), extracted from some of these
results and from some of the $\Upsilon$-decay results discussed in
Sec.~\protect{\ref{sec:upsilon}}.
We note that the LHCb result was obtained with a data sample of
1~fb$^{-1}$, only a third of the currently available sample. 
Analysis of LHCb's 50~fb$^{-1}$ sample with improved triggers, epected
during Run~2 of the LHC, will significantly improve these results.
Older results on these decays by
\babar~\cite{Aubert:2008ps} and Belle~\cite{Wei:2009zv} have been
used to set limits on axion-portal parameters~\cite{Freytsis:2009ct}.

\begin{figure}[!ht]
\begin{center}
\includegraphics[width=1.0\columnwidth]{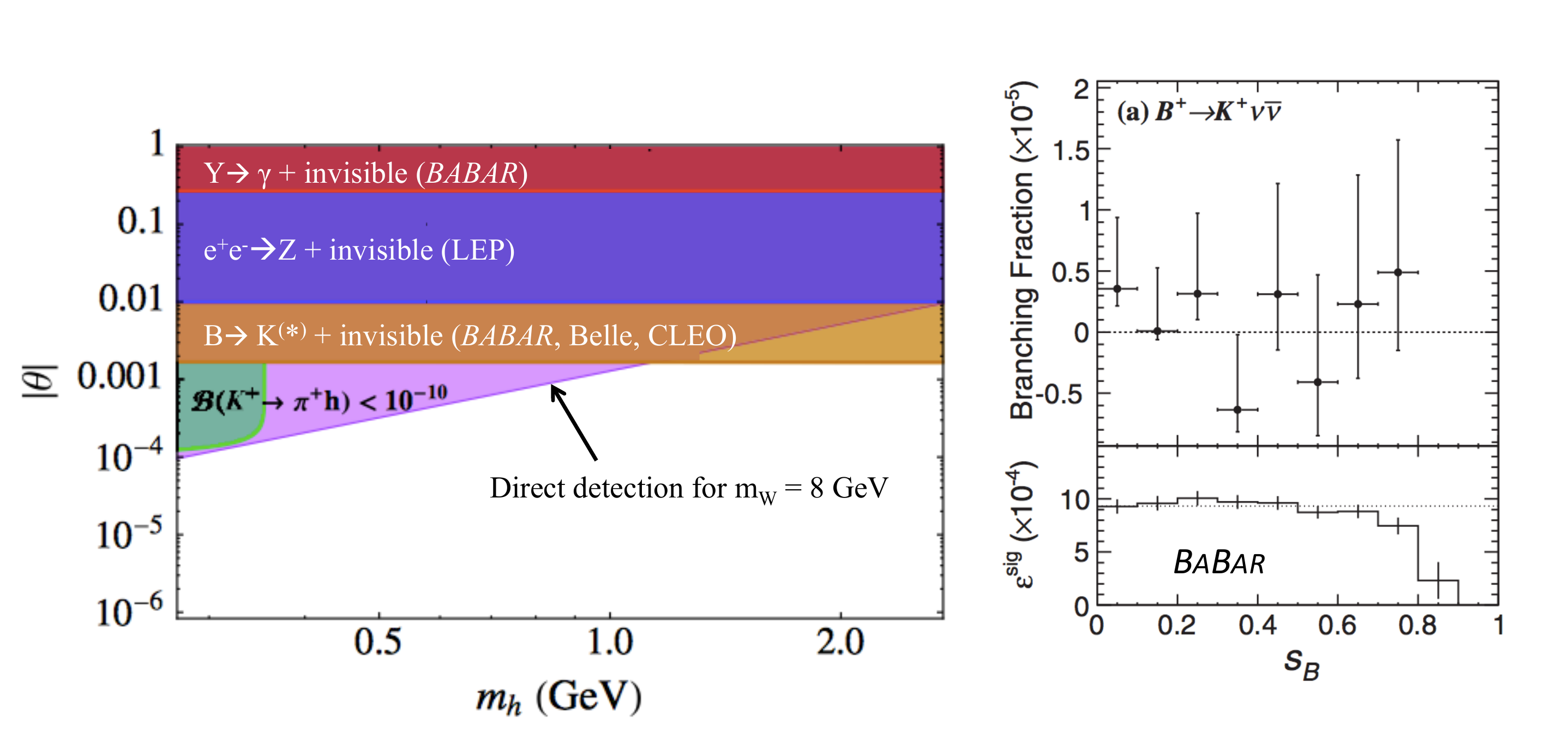}
\caption{\label{fig:BtoKnunu} (a) Central values with uncertainties
  for the partial branching fraction of  $B^+\to K^+\nu\bar\nu$ in bins
  of $s_B = m(\nu\bar\nu)/m_B$ from
  \babar~\protect{\cite{Lees:2013kla}}. The bottom plot shows the
  efficiency variation with $s_B$.  
  (b) Limits~\protect{\cite{Anchordoqui:2013bfa}} on the Higgs-mixing
  parameter $\theta$ of Eq.~(\protect{\ref{eq:theta}}), based on the 
  limits on the total branching fractions for $\Upsilon(1S)\to\gamma 
  + {\rm invisible}$~\protect{\cite{delAmoSanchez:2010ac}},
  $B\to K^{(*)} + {\rm invisible}$~\protect{\cite{Lutz:2013ftz,
      Lees:2013kla,older-babar-knunu,Browder:2000qr}},
  measurements of 
  ${\cal B}(K^+\to \pi^+ + {\rm invisible})$~\protect{\cite{Adler:2001xv,
      Artamonov:2009sz}},
and limits on the cross section for $e^+e^-\to Z^0 + {\rm invisible}$ at 
LEP~\protect{\cite{Barate:1999uc, Abdallah:2003ry, Achard:2004cf, 
    Abbiendi:2007ac}}.
Also shown are the limits from direct-detection experiments, assuming a WIMP
mass of $m_\chi=8$~GeV.
}
\end{center}
\end{figure}

Limits on the branching fractions of $B\to K^{(*)}\phi$ with the
scalar $\phi$ decaying into invisible WIMPS can be extracted from
Belle~\cite{Lutz:2013ftz} and
\babar~\cite{Lees:2013kla,older-babar-knunu} searches for
$B\to K\nu\bar\nu$, as well as a lower-luminosity search by
CLEO~\cite{Browder:2000qr}.  As an example, we show in
Fig.~\ref{fig:BtoKnunu}(a) the results for ${\cal B}(B^+\to
K^+\nu\bar\nu)$ as a function of $s_B = m(\nu\bar\nu)/m_B$.
Fig.~\ref{fig:BtoKnunu}(b) shows limits~\cite{Anchordoqui:2013bfa} on
the absolute value of the mixing parameter $\theta$, defined using
\beq
\tan 2\theta = {g_\theta \left<\phi\right> \left<\phi_{\rm SM}\right> \over
                \lambda_{\rm SM} \left<\phi_{\rm SM}\right>^2 
                -\lambda \left<\phi\right>^2},
\label{eq:theta}
\eeq
where $\phi_{\rm SM}$ is the SM Higgs field, $\lambda_{\rm SM}$
is the usual SM Higgs quartic coupling, and $\lambda$ and $g_\theta$
are defined by the Lagrangian
\beq
{\cal L} = \partial_\mu \phi^\dagger \partial ^\mu \phi 
         + \mu^2 \phi^\dagger\phi - \lambda (\phi^\dagger\phi)^2 
         - g_\theta (\phi^\dagger\phi) (\phi_{\rm SM}^\dagger\phi_{\rm SM})
         + {\cal L}_{\rm SM}.
\eeq
These $|\theta|$ limits are based on the limits on the total (as
opposed to $m(\nu\bar\nu)$-dependent) branching fractions of the
processes shown. Naively, the full Belle-II experiment will produce an
order-of-magnitude improvement in the limits on $B\to
K\chi\bar\chi$. Further improvement should come about by conducting
a dedicated peak search rather than by using just the total branching
fractions.

\section{Summary}

Results from the $B$~factories, fixed-target experiments, and
flavor-physics measurements at hadron colliders have set tight limits
on the parameter spaces of new-physics models involving low-mass
bosons. We have presented limits on parameters of vectors and scalars
that may couple to stable dark-matter particles, together comprising
what has come to be called the dark sector.  Constraints from current
measurements are already ruling out significant regions of
model-parameter space.  Higher sensitivities will be achieved by the
next generation of $B$-factory and fixed-target experiments, as well
as from Run~2 of the LHC.

\bigskip
\section{Acknowledgments}

This research was supported in part by grant No. 2012017
from the United States-Israel Binational Science Foundation (BSF),
and by grant No.  1787/11 from the Israel Science Foundation (ISF).
I thank the IPA organizers, who put together an interesting and broad
agenda and organized a well-run and enjoyable conference, as well as the
conference participants, whose excellent talks and subsequent
conversations taught me a great deal.

%
%

%
%
%
%
 
\end{document}